\documentclass[conference]{IEEEtran}
\IEEEoverridecommandlockouts
\usepackage{cite}
\usepackage{amsmath,amssymb,amsfonts}
\usepackage{algorithmic}
\usepackage[pdftex]{graphicx}
\usepackage{textcomp}
\usepackage{xcolor}
\usepackage{url}
\def\BibTeX{{\rm B\kern-.05em{\sc i\kern-.025em b}\kern-.08em
    T\kern-.1667em\lower.7ex\hbox{E}\kern-.125emX}}

\begin{document}
\title{Scalable Timing Coordination of Bell State Analyzers in Quantum Networks\thanks{This work was supported by JST Moonshot R\&D Program Grant Number JPMJMS226C.}}

\author{\IEEEauthorblockN{Yoshihiro Mori}
\IEEEauthorblockA{\textit{Emerging Media Initiative,} \\
\textit{Kanazawa University}\\
Kanazawa, Japan \\
mori4416@staff.kanazawa-u.ac.jp}
\and
\IEEEauthorblockN{Toshihiko Sasaki}
\IEEEauthorblockA{
\textit{Department of Applied Physics,}\\
\textit{Graduate School of Engineering} \\
\textit{The University of Tokyo}\\
Tokyo, Japan \\
sasaki@qi.t.u-tokyo.ac.jp}
\and
\IEEEauthorblockN{Rikizo Ikuta}
\IEEEauthorblockA{\textit{Graduate School of Engineering Science} \\
\textit{Osaka University}\\
Osaka, Japan \\
ikuta@mp.es.osaka-u.ac.jp}
\and
\IEEEauthorblockN{Kentaro Teramoto}
\IEEEauthorblockA{\textit{Mercari R4D} \\
\textit{Mercari, Inc. }\\
Tokyo, Japan \\
zigen@mercari.com}
\and
\IEEEauthorblockN{Hiroyuki Ohno}
\IEEEauthorblockA{\textit{Emerging Media Initiative,} \\
\textit{Kanazawa University}\\
Kanazawa, Japan \\
hohno@staff.kanazawa-u.ac.jp}
\and
\IEEEauthorblockN{Michal Hajdu\v{s}ek}
\IEEEauthorblockA{
\textit{Graduate School of Media and Governance}\\
\textit{Keio University}\\
Fujisawa, Japan \\
michal@sfc.wide.ad.jp}
\and
\IEEEauthorblockN{Rodney Van Meter}
\IEEEauthorblockA{
\textit{Faculty of Environment and Information Studies} \\
\textit{Keio University}\\
Fujisawa, Japan \\
rdv@sfc.wide.ad.jp}
\and
\IEEEauthorblockN{Shota Nagayama}
\IEEEauthorblockA{
\textit{Mercari R4D, Mercari, Inc.}\\
Tokyo, Japan \\
\textit{Graduate School of Media and Governance, Keio University}\\
Fujisawa, Japan \\
shota@qitf.org}
}

\maketitle

\begin{abstract}
The optical Bell State Analyzer (BSA) plays a key role in the optical generation of entanglement in quantum networks. The optical BSA is effective in controlling the timing of arriving photons to achieve interference. It is unclear whether timing synchronization is possible even in multi-hop and complex large-scale networks, and if so, how efficient it is.  
We investigate the scalability of BSA synchronization mechanisms over multiple hops for quantum networks both with and without memory in each node.
We first focus on the exchange of entanglement between two network nodes via a BSA, especially effective methods of optical path coordination in achieving the simultaneous arrival of photons at the BSA.
In optical memoryless quantum networks, including repeater graph state networks, we see that the quantum optical path coordination works well, though some possible timing coordination mechanisms have effects that cascade to adjacent links and beyond, 
some of which was not going to work well of timing coordination.
We also discuss the effect of quantum memory, given that end-to-end extension of entangled states through multi-node entanglement exchange is essential for the practical application of quantum networks.
Finally, cycles of all-optical links in the network topology are shown to may not be to synchronize, this property should be taken into account when considering synchronization in large networks.
\end{abstract}


\section{Introduction}
\label{sec:intro}

Quantum networks enable distributed quantum computation by distributing entanglement and by transmitting qubits among a set of nodes~\cite{VanMeter2014, Wehner2018, rfc9340}.
Quantum networks contribute to large-scale quantum computers, such as fault-tolerant quantum computers capable of executing quantum algorithms with quantum advantage~\cite{VanMeter2006,kim09:_integ_optic_ion_trap,oi06:_dist-ion-trap-qec,jiang2010scalable,ahsan2015designing},
and the quantum Internet to utilize quantum phenomena in wide-area distributed quantum computation, such as secure communication and computation, rapid leader election, byzantine agreement, sensing such as ultra-long baseline interferometry for telescopes, and spatial and temporal reference frame synchronization~\cite{Kimble2008-eo, Cirac1999, Bennett1984, OBrien2014, Gottesman2012, Wang2023usecase, Broadbent2009}.

The loss of photons in quantum networking is critical because, unlike in classical networking, we cannot copy quantum data for backup due to the no-cloning theorem~\cite{Wootters1982no-cloning, Park1970no-cloning}. 
Since we can repeat attempts to create entanglement but cannot afford to lose critical quantum data, the consensus approach today is to build applications on top of quantum network architectures that provide end-to-end entanglement generation as the fundamental service~\cite{VanMeter2021}.
Setting aside the crucial issue of errors, first, quantum links generate raw Bell pairs between adjacent nodes; second, such Bell pairs are converted to end-to-end Bell pairs with a dedicated operation called entanglement swapping~\cite{Briegel1998}; finally, the end nodes use the end-to-end Bell pairs for quantum teleportation or other distributed applications.

Early quantum link architecture proposals assume the existence of quantum memories that are used both in the process of generating raw Bell pairs and for performing entanglement swapping.
Memory-Interference-Memory (MIM) and Memory-Memory (MM) use emission of photons from quantum memories to generate raw Bell pairs, while the Memory-Source-Memory (MSM) architecture uses an entangled photon pair source (EPPS) node, which generates raw Bell pairs as photon pairs, with quantum memory nodes.
A common component shared by these architectures is the optical Bell State Analyzer (BSA), which is used to herald successful generation of link-level Bell pairs.
Single memory-based quantum links have been implemented in pioneering experimental demonstrations~\cite{Yang2006, Sun2017, Krutyanskiy2023, Pu2021, Sun2017-1, krutyanskiy2023telecom,lu2022micius}.
However, it is not yet clear whether these architectures will lead to near-term demonstrations of quantum networks due to the tremendous technical challenges of deploying quantum memories in real-world environment.

It is therefore believed that memory-less quantum 
repeaters offer a more suitable path to near-term deployment of quantum networks.
Two quantum link architectures without quantum memories are under development, with advances expected from probabilistic toward deterministic entanglement generation. The first form utilizes chains of entanglement sources (Source, or S) and BSAs (Interference, or I).
We can incorporate BSAs and additional sources into a path, giving us optical paths such as DSD, DSISD, DSISISD, etc., where 'D' denotes a detector.
Any such path requires synchronization of the photon arrivals at each BSA, creating timing dependencies between the nodes. 
We can refer to such paths or sets of nodes (which may be optically switched) as ``photonic synchronization domains'', or PSDs (similar in spirit, if not technology, to coaxial Ethernet's collision domains, denoting particularly timing-constrained sets of links and nodes).
The current state of the art in photonic Bell pair generation may have an efficiency of only $0.1\% \sim 1\%$ per trial~\footnote{With technologies such as spontaneous parametric down conversion (SPDC), the probability can be raised by increasing pump laser power within certain limits, but at the expense of fidelity.}. With such technologies, the scalability of a PSD to multi-hop optical paths will be limited by the simultaneous Bell pair generation and detection probability, which decays exponentially in the number of hops. Nevertheless, we expect small numbers of hops to be useful, and such paths are the focus of this paper.

The other memory-less link architecture is known as all-photonic quantum repeaters, a future approach that utilizes repeater graph states to overcome photon loss and faulty measurement devices, becoming nearly deterministic in end-to-end entanglement generation~\cite{azuma2015all}.
Much of the current research focuses on efficient generation of these repeater graph states~\cite{hilaire2021resource, Sun2022} and end node participation~\cite{benchasattabuse2023architecture}.
Coordinated photon arrival at intermediate measurement nodes is a crucial yet underexplored element of all-photonic link architectures.

In this paper, we investigate the scalability of coordinated photon emission and arrival that would lead to successful entanglement swapping at the intermediate BSA support nodes.
The primary question that we ask is whether the required control negotiation can be confined to the individual links or whether it propagates to adjacent links and possibly beyond.

\begin{figure}[htbp]
\centering
\includegraphics[width=\columnwidth]{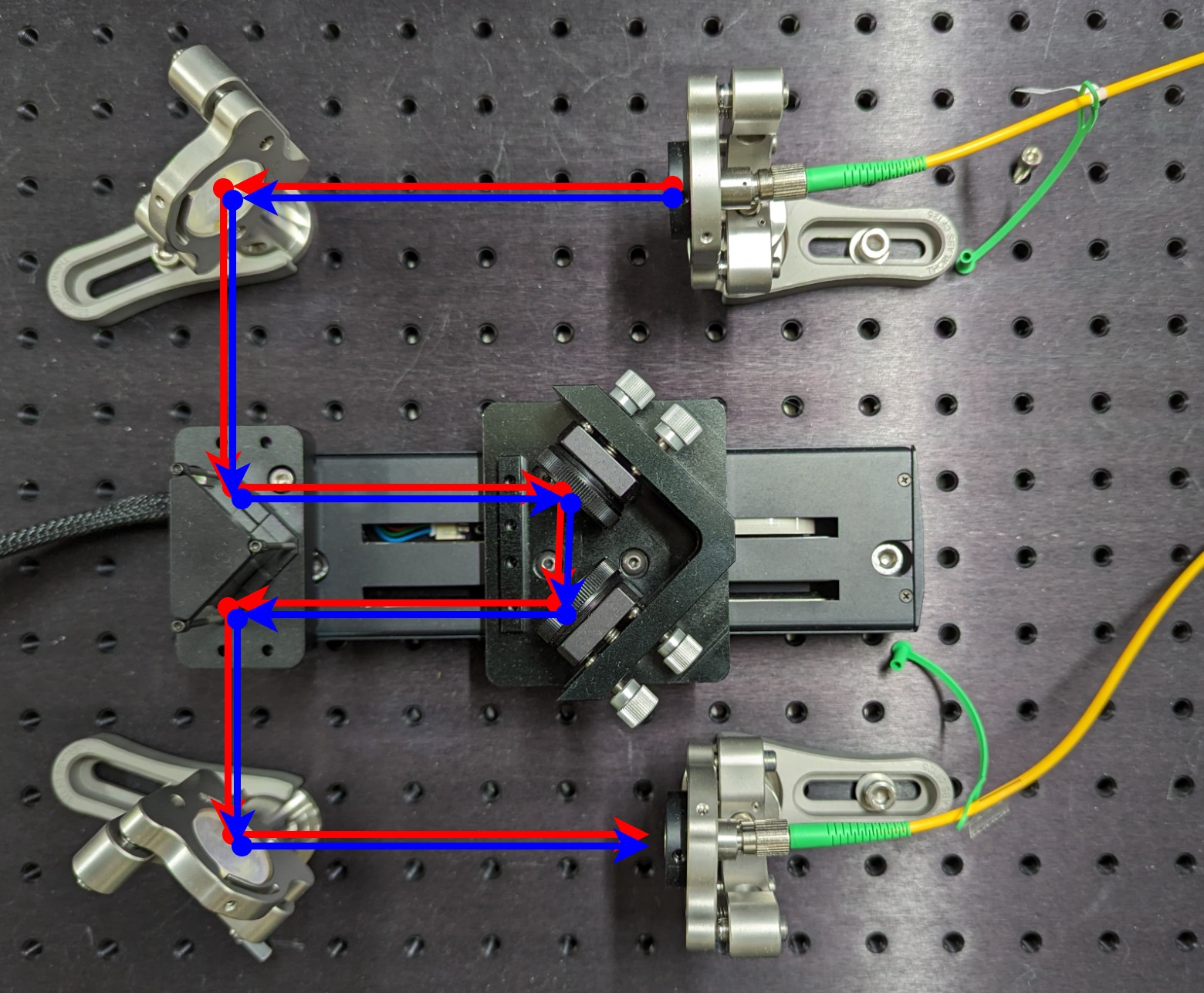}
\caption{An optical delay line (ODL) is a common tool for adjusting the arrival time of single photons or laser pulses in a simple configuration; the slide (center) can be moved under programmatic control to lengthen or shorten the path from one fiber (yellow) to the other. Using multiple such devices in a single photonic synchronization domain is a challenge. The red and blue arrows represent the optical paths of classical control and quantum channels.
Details are given in the text.}
\label{fig:odl}
\end{figure}

Our result shows that the mechanisms adjusting the timing of photons in the quantum channel work scalably. 
Such adjustment can be achieved with such as optical delay lines (See Fig.~\ref{fig:odl}).
In quantum networks without quantum memories, several nodes along a path each generate two photons in the Bell state and transmit them to their respective neighboring optical BSAs.  If we apply the timing adjustment in the quantum channel after the emission of the photons (rather than applying it in the classical control channel or adjusting photon emission timing), negotiation and synchronization can be confined to each individual link without propagating to adjacent links and beyond. We find that the BSA support node is the correct locus for control of the timing. When introducing long-lived quantum memories, negotiation and synchronization do not propagate to adjacent links because quantum memory decouples the timing of link-level Bell pair generation from inter-link entanglement swapping. 
Finally, we discussed the potential extension of this mechanism to network topologies with cycles.

Our investigation offers a possible route towards practical implementation of multi-hop quantum networks.
    
\section{Entanglement Swapping and the Optical BSA}
\label{sec:es}

In the BSA, a crucial role in photon entanglement is played by Hong-Ou-Mandel (HOM) interference\cite{HOM}. 
HOM interference exploits the phenomenon where two indistinguishable photons entering a balanced beam splitter simultaneously exit through the same port. 
This interference phenomenon leads to erasure of which-path information, a crucial step in optical Bell-state measurement \cite{Bouwmeester1997-qk}.

In order to maximize the fidelity of end-to-end Bell pairs, it is crucial that the photons that are measured at the BSA are indistinguishable in all of their degrees of freedom.
This includes not only their arrival time, but also their spectral modes, polarization, and temporal modes \cite{branczyk2017hong}.
Our proposed BSA synchronization mechanism focuses primarily on ensuring the same arrival time of the photons at the BSAs.

To begin the discussion, we set up an ideal experimental system. 
First, Bell pair photons are continuously generated at a node within the network, serving as the medium for quantum state transmission. 
Within the network nodes, the generation of entangled photons employs the process known as SPDC (Spontaneous Parametric Down-Conversion)~\cite{kwiat95:PhysRevLett.75.4337}. In this process, the pump light enters a nonlinear optical crystal. Due to the crystal's nonlinearity, a single pump photon is converted into two lower-energy, entangled photons. 
Multiple such network nodes are prepared, and support nodes with BSAs are placed between them. 
From each network node, one of Bell pair photons is sent through a path such as an optical fiber to the support node. 
Assuming entanglement swapping occurs at the BSA, this results in the generation of quantum entanglement across the entire network.

\begin{figure}[htbp]
\centering
\includegraphics[width=\columnwidth]{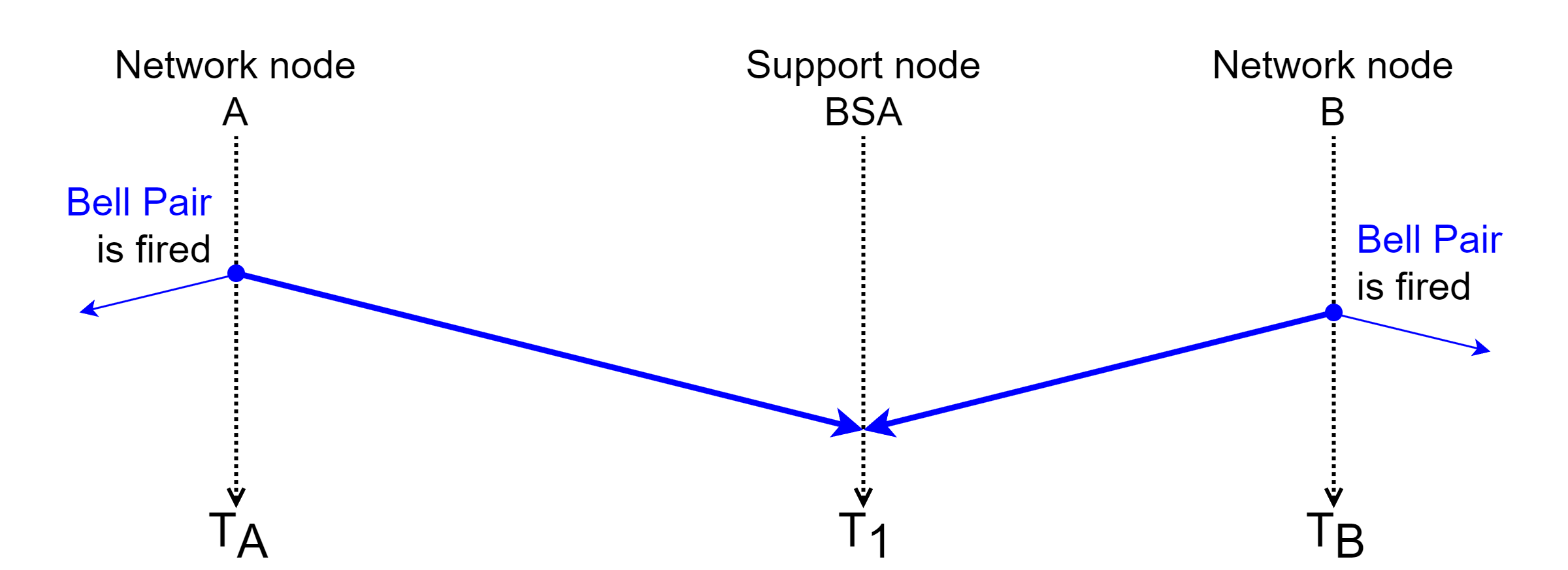}
\caption{Schematic diagram of a quantum network for realizing entanglement exchange between two network nodes using support nodes (BSA)}
\label{fig:fig1}
\end{figure}

A schematic of this is shown in Fig. \ref{fig:fig1}. 
Here, the horizontal spread represents spatial extension, indicating that network nodes A and B are different devices placed at different locations. 
A support node with BSA is installed between the two network nodes. 
The vertical lines represent the passage of time. 
At a certain moment, one photon of a Bell pair generated at network node A is sent through an optical fiber or similar channel to the BSA support node. 
As the one of Bell pair photon approaches the BSA over time, this quantum channel is depicted by a rightward downward diagonal blue arrow. 
Similarly, for the one of Bell pair photon generated at node B, a leftward downward diagonal blue arrow is depicted. 
If these two Bell pair photons arrive at the BSA simultaneously, interference occurs, and entanglement swapping takes place, projecting the two photons into a single Bell pair. 
The other photons generated at A and B become entangled, extending entanglement end-to-end.

In Fig. \ref{fig:fig1}, there is no timing discrepancy 
between photons arriving at the BSA; however, in reality, timing discrepancies arise due to differences in the length of optical fibers (as support nodes are not always equidistant from network nodes), as well as changes in length due to temperature variations and fluctuations caused by natural/artificial factors.
We denote this discrepancy as $\Delta$. 
In the model using the BSA as a support node, a strategy to correct this $\Delta$ is necessary.

\section{Synchronization Models for the Optical BSA Between Two network nodes}
\label{sec:2NN}

In constructing a quantum network, for scalability, we examine the model of entanglement swapping performed by BSA between two network nodes. 
The key to this process is the simultaneous arrival of the two photons at the BSA. 
This timing adjustment relies on the generation of a heralding signal that allows us to measure the value of $\Delta$. 
The BSA is the only location where this measurement can be done accurately, \emph{leading us to choose the BSA as the locus of control for timing synchronization}. The full details of this measurement process and the rate of change over time of $\Delta$ are beyond the scope of this paper.
In order to achieve the simultaneous arrival, the following four strategies can be considered:

\begin{figure}[t]
\centering
\includegraphics[width=\columnwidth]{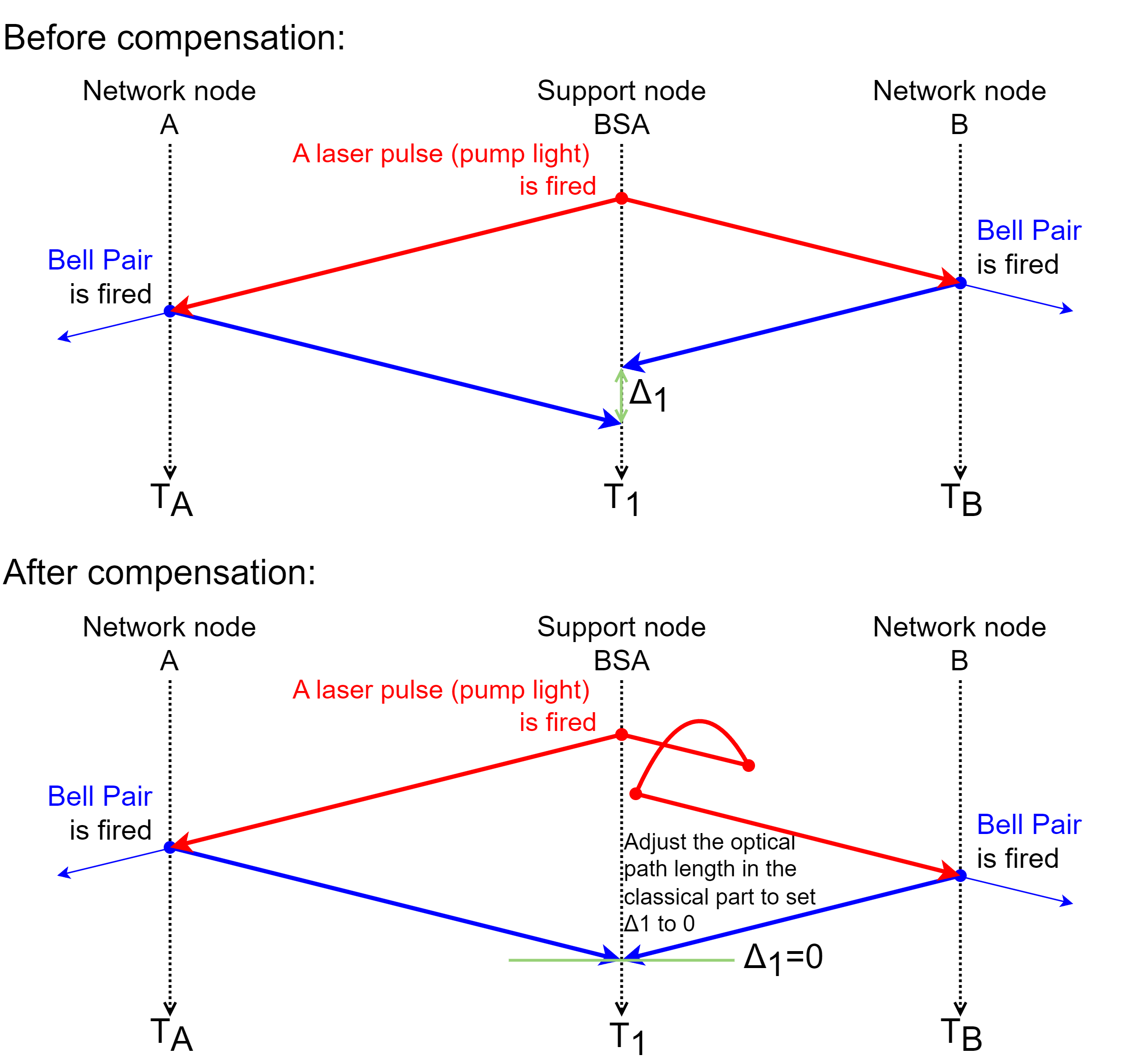}
\caption{A laser pulse (pump light)  is fired from the support node, generating Bell pair photons at the network nodes. One photon from each Bell pair at the network nodes is then fired towards the support node, causing entanglement swapping at the BSA. However, naively, a time discrepancy $\Delta$ occurs, so adjusting the path of the laser pulses can reduce $\Delta$ to zero.
}
\label{fig:fig2-1}
\end{figure}

1) as shown in Fig. \ref{fig:fig2-1}, the BSA support node directly emits laser pulses (pump light) intended to generate Bell pairs towards network nodes A and B.
Alternatively, it sends a command signal to fire the laser pulses, which, upon reception, prompts the network nodes to immediately emit laser pulses within the nodes, thus generating Bell pair photons. 
These photons are then sent from network nodes A and B to the BSA support node.
In the figures, the pump pulse (or trigger) is indicated by red arrows (classical control channel) and the single photons are indicated by blue arrows (quantum channel).
At the BSA support node, photons arriving from network nodes A and B are made to interfere to facilitate entanglement swapping, resulting in the photons remaining at network nodes A and B being placed in a quantum entangled state.
There is a timing discrepancy $\Delta$ in the entangled state photons arriving at the BSA support node. 
To compensate this $\Delta$, an optical delay line (ODL, Fig.~\ref{fig:odl}) is introduced in the path of the laser pulse or firing command signal, and the compensation of $\Delta$ within the BSA support node is performed as shown in Fig. \ref{fig:fig2-1}.

\begin{figure}[htbp]
\centering
\includegraphics[width=\columnwidth]{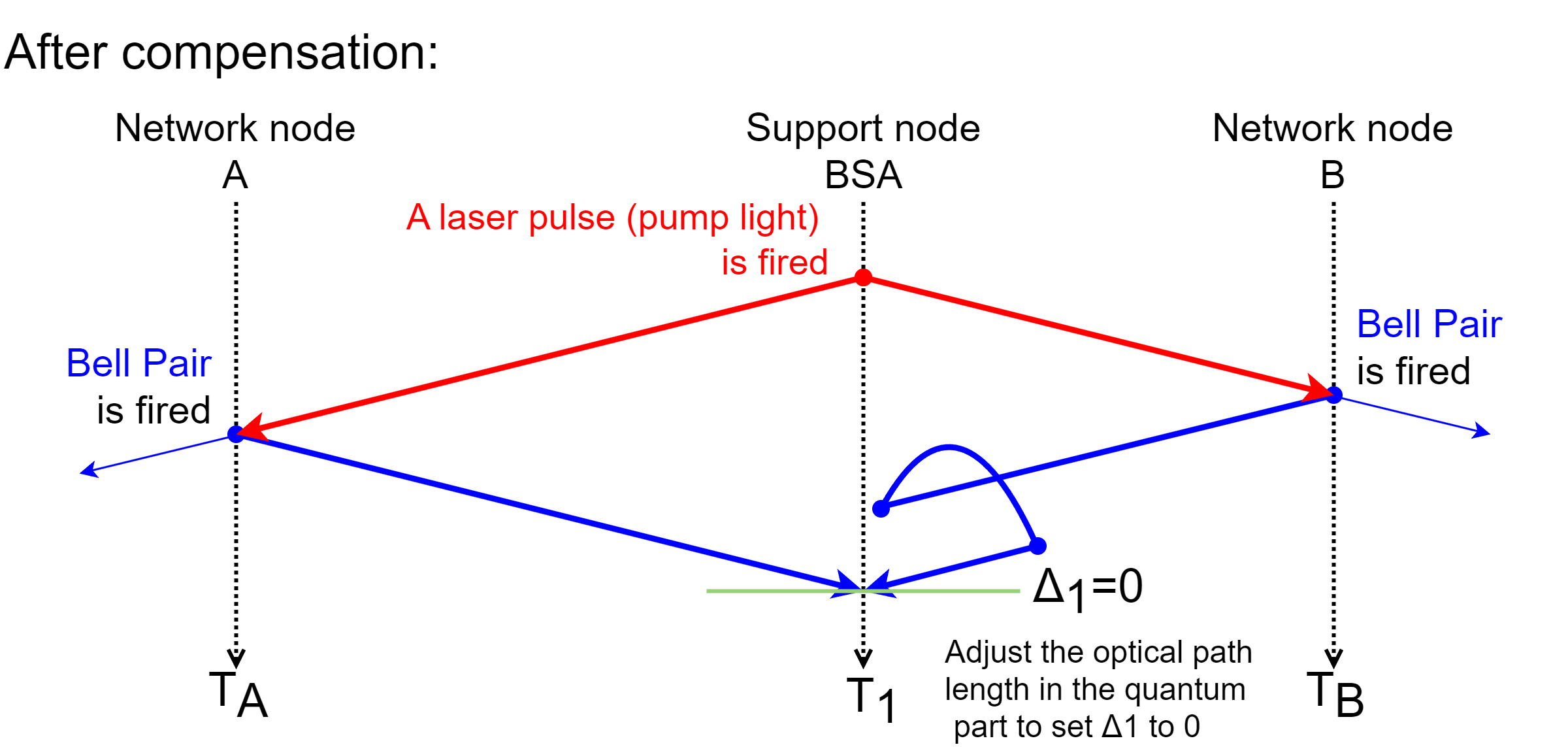}
\caption{One photon from each Bell pair at the network nodes is directed towards the support node to initiate entanglement swapping at the BSA; the optical path of Bell pair photons is compensatedto to reduce $\Delta$ to zero.
}
\label{fig:fig2-2}
\end{figure}

2) as shown in Fig. \ref{fig:fig2-2}, the BSA support node actively measures the firing timing. 
To compensate this $\Delta$, 2) involves adjusting the optical path that transmits the Bell pair photons. 
Here, the adjustment is done in the quantum channel rather than the classical control channel, allowing for the compensation of $\Delta$ within the BSA support node, as demonstrated in Fig. \ref{fig:fig2-2}.  The use of a quantum buffer memory, described in Sec.~\ref{sec:memory}, can be viewed as a variant of this approach.

\begin{figure}[htbp]
\centering
\includegraphics[width=\columnwidth]{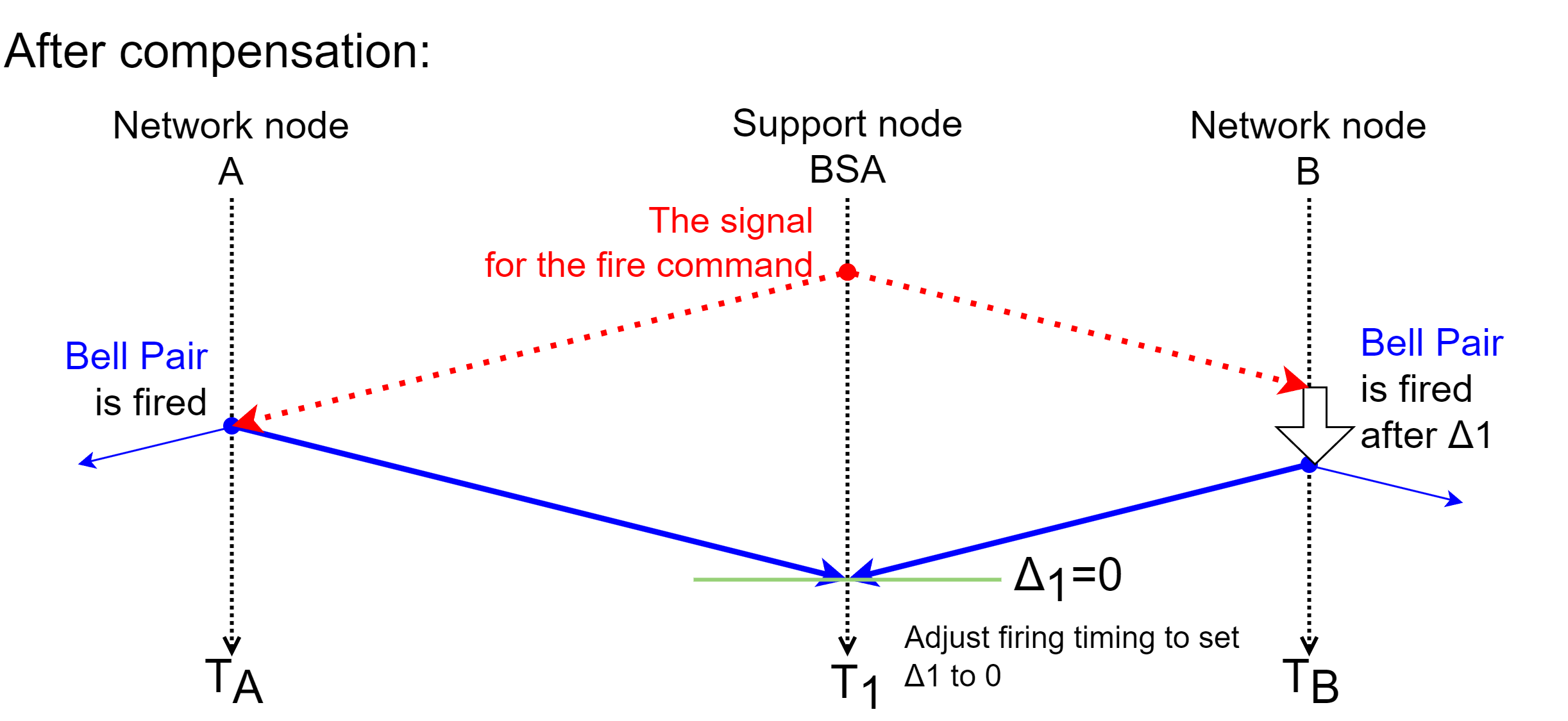}
\caption{The BSA actively measures the firing timing. The command for measuring the firing timing is represented by a red dashed arrow. By staggering the timing at which one photon from each Bell pair is fired from the source to the BSA, $\Delta$ is reduced to zero.
}
\label{fig:fig2-3}
\end{figure}

3) as shown in Fig. \ref{fig:fig2-3}, the BSA support node actively measures the firing timing. 
To compensate this $\Delta$ to 0, 3) involves adjusting the timing of firing the Bell pair photons by adding an offset, allowing for the adjustment of $\Delta$ within the BSA support node, as shown in Fig. \ref{fig:fig2-3}. 
This strategy does not require instructions from the BSA support node to the other nodes.

\begin{figure}[htbp]
\centering
\includegraphics[width=\columnwidth]{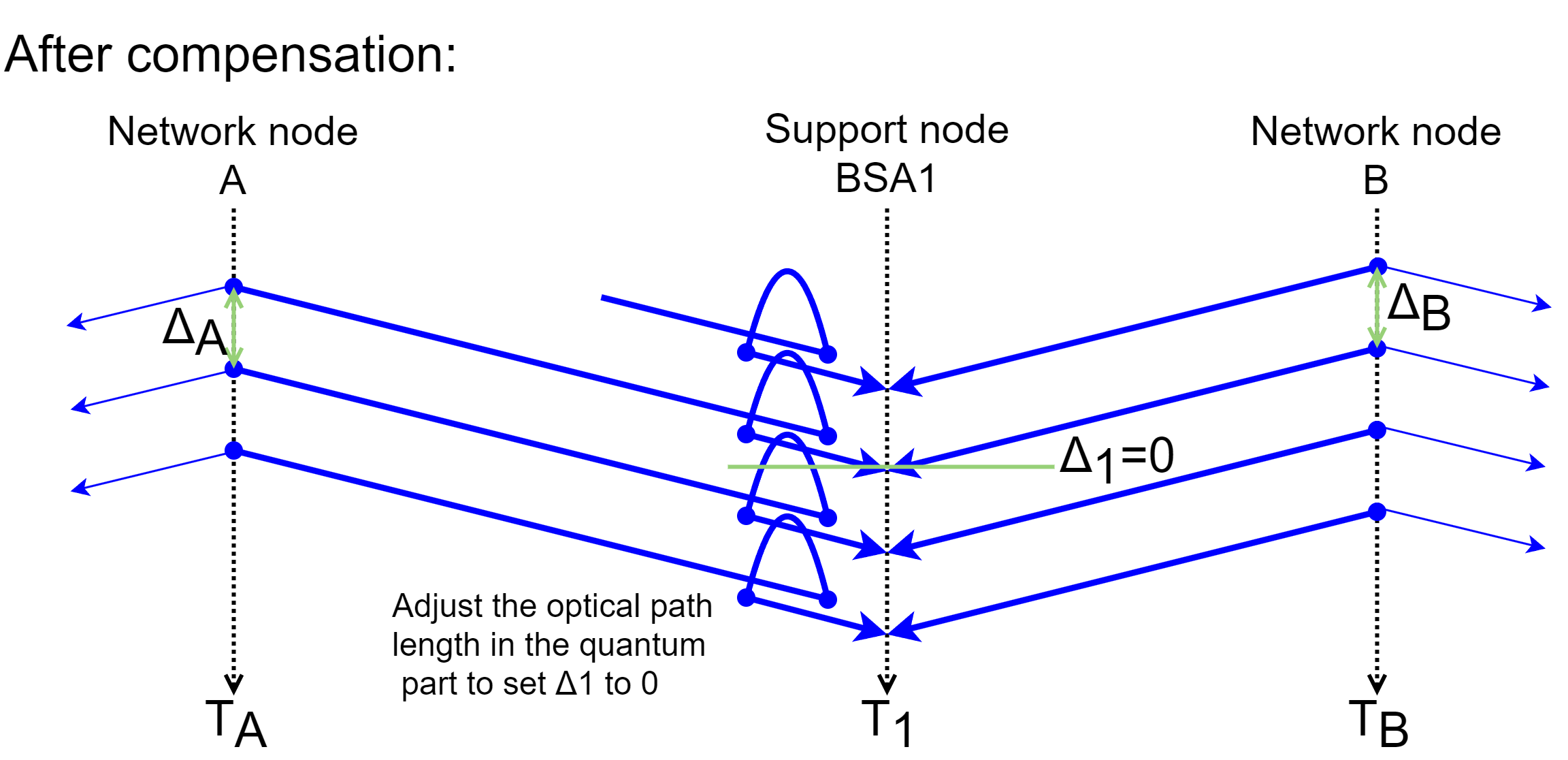}
\caption{Photon generation frequency is synchronized between network nodes directly, so $\Delta_{A}=\Delta_{B}$. The simultaneous arrival is adjusted in the BSA support node.
}
\label{fig:frequency-synchronized-quantum-adjustment}
\end{figure}

4) as shown in Fig. \ref{fig:frequency-synchronized-quantum-adjustment}, 
entangled photon generation obeys not the timing provided by the BSA support node but the frequency beat synchronized directly between nodes A and B. The simultaneous photons' arrival at the BSA is ensured by adjusting the length of the quantum channel.

In practice, installing ODL in the quantum channel should be avoided if possible, 
because adding optical components in quantum channel causes photon losses and decreases the throughput.
Therefore, 1) and 3) are preferable for a single link.

\section{Scalability Issues Among Multiple Nodes}

\begin{figure*}[ht]
\centering
\includegraphics[width=170mm]{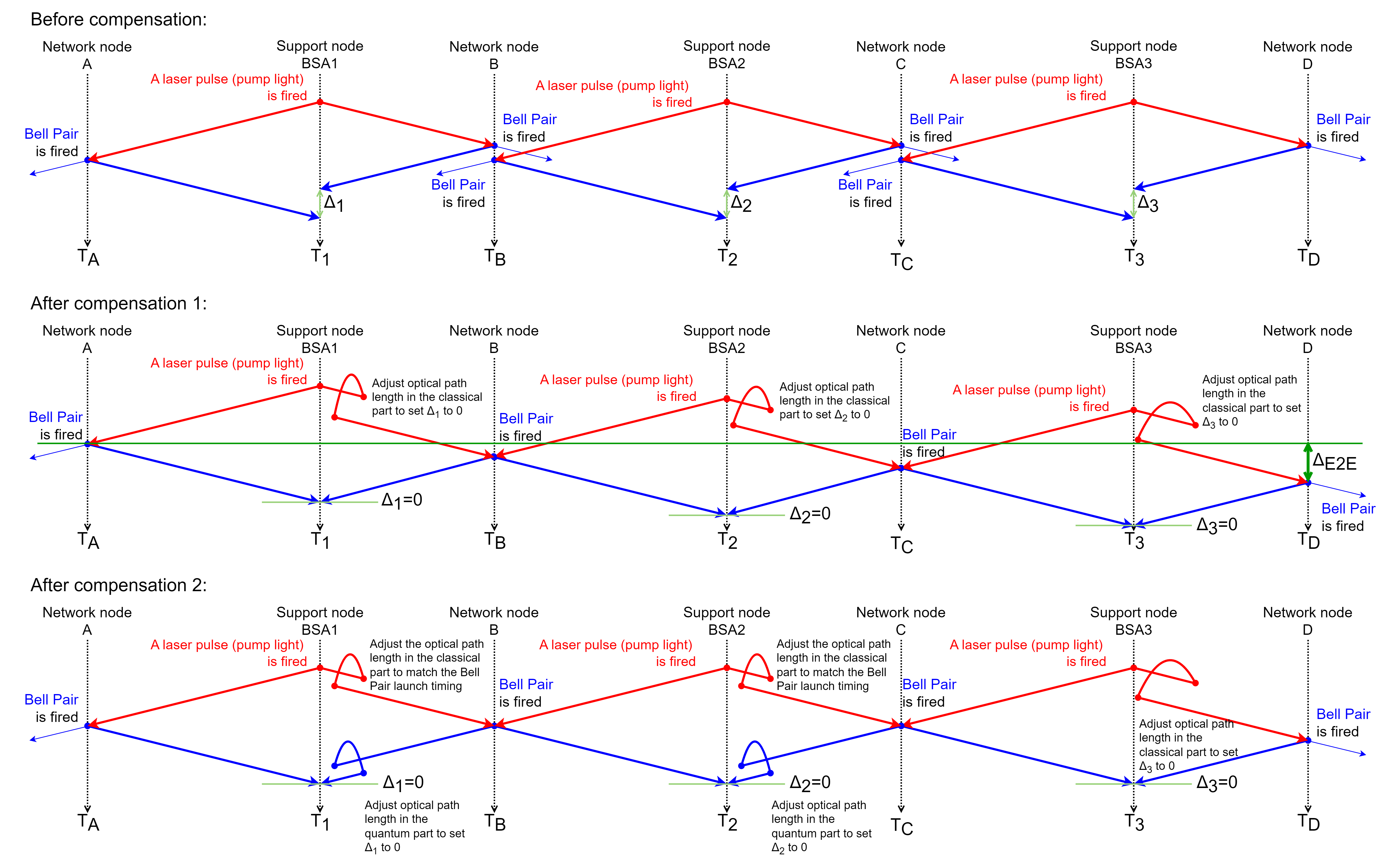}
\caption{Even in the case of four network nodes, each support node either fires a laser pulse to the adjacent network node to generate Bell pair photons or issues an instruction to that network node to generate Bell pair photons (represented by red arrows). Consequently, Bell pair photons are fired from each network node towards the support node (represented by blue arrows). To synchronize the firing timing, adjustments to the classical light path are made. 
If only classical optical path alignment is used, the synchronization information of BSA1 will propagate and change the timing of the firing of the Bell pair photons (After compensation1).
Furthermore, the timing is aligned by compensating the optical path of the entangled photons generated at the network nodes until they reach the BSA (After compensation2).
}
\label{fig:fig3-1}
\end{figure*}

\begin{figure*}[ht]
\centering
\includegraphics[width=180mm]{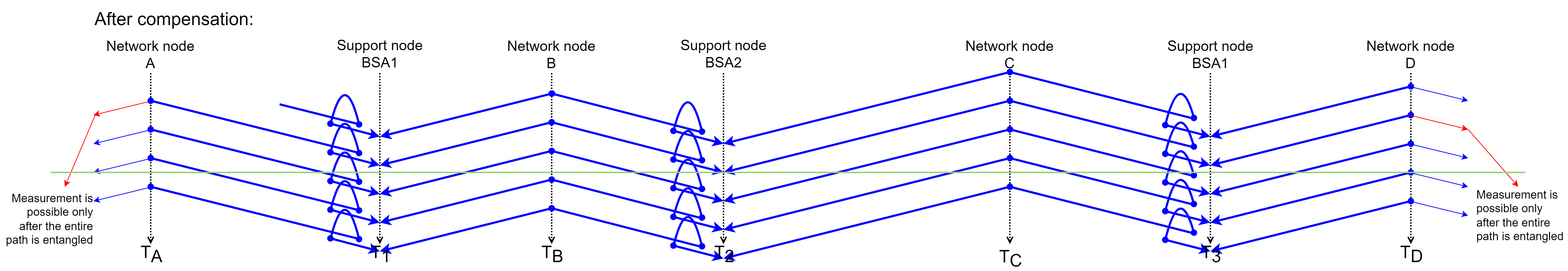}
\caption{Each network node continuously outputs Bell pair photons at fixed timings, and these photons are then directed to the BSA at the support node, where optical path adjustments are made to align the timings, and entanglement swapping is performed as appropriate.
}
\label{fig:fig3-2}
\end{figure*}

So far, we have shown the case with two network nodes and a support node, but here we will demonstrate the case with four network nodes and three BSAs installed. 
As shown in Fig. \ref{fig:fig3-1}, four network nodes from A to D are placed, with support nodes BSA1 to 3 in between. 
First let's consider fed through 1) from Sec.~\ref{sec:2NN} applied to multi-hop networks. 
With 1), it may not be possible to always compensate the timings to fire entangled Bell pair photons from each network node to cause entanglement swapping at the both BSAs; in Fig. \ref{fig:fig3-1}'s "After compensation1", 
a synchronization cascade solves this problem but such a cascade expanding the PSD beyond the adjacent neighbors, therefore such cascade is unacceptable.
In Fig. \ref{fig:fig3-1}'s "After compensation2", each support node either fires a laser pulse to the adjacent network node to generate Bell pair photons or issues an instruction to that network node to generate Bell pair photons. 
In Fig. \ref{fig:fig3-1}, to synchronize the firing timing, adjustments to the classical light path, represented by red arrows, are made.
However, by this adjustment, the timing to reach to a BSA may shift from the original timing the BSA intended.
Therfore, the timing has to be synchronized by adjusting the optical path of the entangled photons on the way to the BSA anyway (The combination of 1) and 2).)
In these case, even if the timing of photon emission for Bell pair by the support node is coordinated, the required control negotiation cannot be confined to individual links but propagates to adjacent links and further to the next adjacent links. 
Because, the effect on the timing to fire the entangled Bell pair photons due to propagation when the number of multi-hops is increased lasts from the beginning to the end.
However, what is desirable is for timing adjustments to be made autonomously at each node.
This reveals that the only necessary timing adjustment at the support nodes is the optical path adjustment of the entangled photons generated at the network nodes.

Therefore, to construct a quantum network using the BSA model, as shown in Fig. \ref{fig:fig3-2}, it is reasonable to continuously output Bell pair photons from each network node at a fixed timing, adjust the timing by optical path adjustments at the support node's BSA, and appropriately perform entanglement swapping; 
therefore the expansion of 4).

The third case, applying just 3), (adjusting the emission timing by adding an offset at the nodes to the timing provided by BSAs), to multi-hop networks also has the problem of synchronization cascades. If node B adjusts the timing for BSA1, that adjustment may not work for BSA2. Then BSA2 has to 
accele to that adjustment, and consequently BSA2 has to negotiate with node.
The synchronization propagates beyond the adjacent node and PSD is not confined into a link.

When implementing a quantum network using the BSA model, the issues become determining the level of timing synchronization accuracy required at all network nodes and the frequency at which Bell pair photons can be generated.
In this issue, the interval of timing synchronization among multiple network nodes and the number of times entangled Bell pair photons can be generated during that interval are important.
This is because if entangled Bell pair photons can be generated at the shortest possible interval, timing synchronization of entangled exchanges at shorter intervals is possible.

\section{Extension with the Use of Quantum Memory}
\label{sec:memory}

In addition to this, in quantum networks using the BSA synchronization mechanism, there is also the issue of how to maintain one of the Bell pair photons generated at the end node until entanglement swapping is completed across the entire network. 
The introduction of quantum memory can be considered as a means to solve this problem.

The link architectures introduced in Sec.~\ref{sec:intro} (MM, MIM and MSM), as proposed in the earliest literature on quantum repeaters, involve the use of quantum memory. A quantum memory is a device capable of temporarily storing quantum information and retrieving it later in a reusable form~\cite{Simon2010-cn}. 
Various physical systems, including atoms, ions, or superconducting qubits, are utilized for quantum memory. 
The performance of quantum memory is evaluated based on the number of qubits it can store, the coherence time (the duration over which quantum information is preserved), and the efficiency of read and write operations. Some memories allow computation on the qubits they store, while others simply buffer quantum data.
The development of quantum memory, capable of storing quantum states for several milliseconds and retrieving them with high fidelity, is essential for quantum repeater technology that enables long-distance transmission of quantum information.

In the case of quantum memory that stores a quantum state as an optical mode and then releases it again~\cite{buser22:storage}, the argument up to this point can be used directly.
From the timing control point of view, this buffering of an optical mode is the same as the insertion of a delay in the communication channel.
The results so far show that the asymmetry in the length of the communication channel is not a problem in terms of timing control.
Therefore, the key requirement for quantum memory is the ability to preserve quantum states for the duration needed for entanglement swapping and the transmission of classical information.

If the quantum memory supports long and variable memory buffering time, more things can be done.
It makes possible to perform an entanglement swap between entangled quantum states that were generated at such widely different times.
It cannot be handled only with optical path adjustment (Fig.~\ref{fig:odl}).
This can increase the probability of end-to-end quantum entanglement swap success.

Memory also provides advantages in terms of network resource control.
In the present paper, it is assumed that the timing of when to use the communication channel is to be allocated synchronously.
If the quantum memory time is long and variable, the communication channel can be allocated in the hop-by-hop manner, as in packet switching in the classical Internet.
This is important to simplify network resource control and to prevent network control from becoming a bottleneck when the network is expanded. Thus, we can see a clear progression of increasing capabilities and performance as newer technologies are introduced.

\section{Cycles and Multiple Paths}

Advances in quantum network technology are expected to enable the development of more efficient light sources and the construction of quantum links capable of supporting a larger number of hops in the future.

\begin{figure}[htbp]
\centering
\includegraphics[width=\columnwidth]{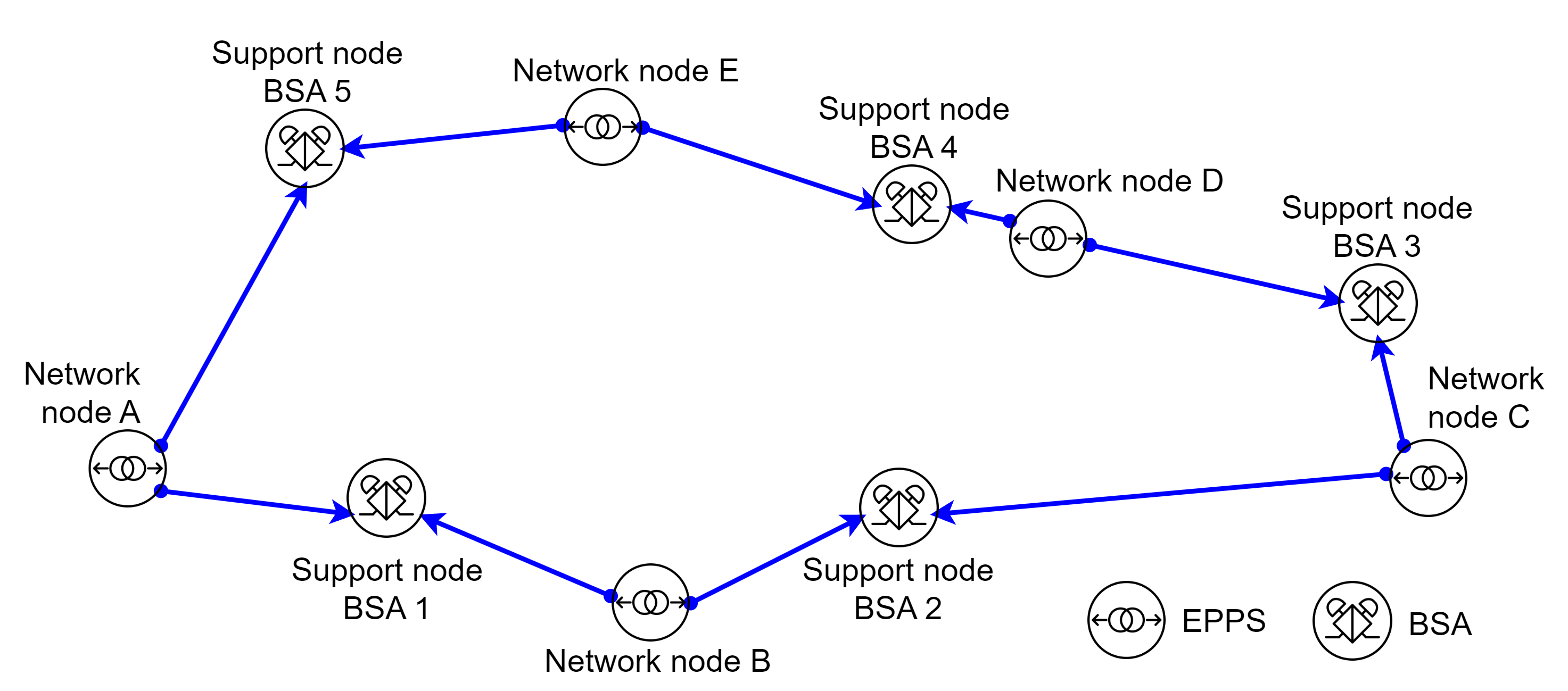}
\caption{
When an entanglement is to be created between network nodes A and C in this network, two paths through Node B or D and E exist. The blue arrows schematically represent spatial optical paths of the entangled photons. In this network, it may not be possible to synchronize timing of outgoing or incoming photons in all nodes and BSA simultaneously if the range of optical path adjustment is limited.
}
\label{fig:fig4}
\end{figure}

In this case, multiple paths as in Fig.~\ref{fig:fig4} can be used at the same time to distribute entanglement.
It can enhance redundancy and fault tolerance, improve the efficiency of quantum entanglement distribution, and facilitate applications in quantum error correction.
From the viewpoint of timing synchronization, entanglements in the multiple paths may not be delivered simultaneously if the range of optical path adjustment is limited. 
It is impossible to synchronize the timing of entanglement distribution of a cyclic path in general. 
This synchronization problem may result in problems to generate multi-partite quantum state among multiple nodes.

\section{Conclusion}

In this paper, we investigated the necessity of and methods for extending optical entanglement swapping end-to-end in the construction of quantum networks.
Particularly, the examination of expanding a photonic synchronization domain from two network nodes to multiple nodes is of importance for near-term experiments as well as long-term all-photonic designs.
The results of this study demonstrate that having the BSA drive optical path adjustment, ensuring the simultaneous arrival of Bell pair photons generated by network nodes, is the most feasible method for effectively achieving entanglement swapping at the BSA. 
Fortunately, we conclude that even within a PSD the need for synchronization can be limited in scope by applying timing adjustments in the quantum channel at the BSA.
As the technology advances, introducing quantum memory will define a boundary for the PSD, improving the efficiency and flexibility of entanglement swapping. 
These findings provide important guidelines for the construction of long-distance quantum networks.

\section*{Acknowledgment}
DeepL was used to assist in translation.

\bibliographystyle{unsrt}
\bibliographystyle{IEEEtran.bst}
\bibliography{references}

\end{document}